\documentclass[twocolumn]{bmcart}

\usepackage{amsthm,amsmath,amssymb}
\usepackage[utf8]{inputenc}
\usepackage{url,graphicx} 

\startlocaldefs
\newcommand{\ignore}[1]{}
\endlocaldefs

\begin{document}
\begin{frontmatter}
\begin{fmbox}
\dochead{Research}
\title{Phylogenetic Convolutional Neural Networks in Metagenomics}
\author[
   addressref={aff1},
   noteref={first},
   email={fioravanti.diego@gmail.com}
 ]{\inits{DF}\fnm{Diego} \snm{Fioravanti}}
\author[
   addressref={newaff1}, 
   noteref={first},
   email={ylenia.giarratano@gmail.com}   
]{\inits{YG}\fnm{Ylenia} \snm{Giarratano}}
\author[
   addressref={aff1},
   noteref={first},
   email={vmaggio@fbk.eu}
]{\inits{VM}\fnm{Valerio} \snm{Maggio}}
\author[
    addressref={aff2},
    email={claudio.agostinelli@unitn.it}
 ]{\inits{CA}\fnm{Claudio} \snm{Agostinelli}}
\author[
   addressref={aff1},
   email={chierici@fbk.eu}
]{\inits{MC}\fnm{Marco} \snm{Chierici}}
\author[
   addressref={aff1},
   corref={aff1},
   email={jurman@fbk.eu}
]{\inits{GJ}\fnm{Giuseppe} \snm{Jurman}}
 \author[
   addressref={aff1},
   email={furlan@fbk.eu}
  ]{\inits{CF}\fnm{Cesare} \snm{Furlanello}}

\address[id=aff1]{%
  \orgname{Fondazione Bruno Kessler (FBK)},
  \street{Via Sommarive 18 Povo},
  \postcode{I-38123}
  \city{Trento},
  \cny{Italy}
}
\address[id=newaff1]{%
  \orgname{Centre for Medical Informatics, Usher Institute, University of Edinburgh},
  \street{Teviot Place},
  \postcode{EH8 9AG},
  \city{Edinburgh},
  \cny(UK)
}
 \address[id=aff2]{%
   \orgname{Department of Mathematics, University of Trento},
   \street{Via Sommarive 14 Povo},
   \postcode{I-38123}
   \city{Trento},
   \cny{Italy}
 }

\begin{artnotes}
\note[id=first]{Joint first author}     
\end{artnotes}

\begin{abstractbox}

\begin{abstract} 
\parttitle{Background} 
Convolutional Neural Networks can be effectively used only when data are endowed with an intrinsic concept of neighbourhood in the input space, as is the case of pixels in images. 
We introduce here Ph-CNN, a novel deep learning architecture for the classification of metagenomics data based on the Convolutional Neural Networks, with the patristic distance defined on the phylogenetic tree being used as the proximity measure. 
The patristic distance between variables is used together with a sparsified version of MultiDimensional Scaling to embed the phylogenetic tree in a Euclidean space. 

\parttitle{Results} 
Ph-CNN is tested with a domain adaptation approach on synthetic data and on a metagenomics collection of gut microbiota of 38 healthy subjects and 222 Inflammatory Bowel Disease patients, divided in 6 subclasses. Classification performance is promising when compared to classical algorithms like Support Vector Machines and Random Forest and a baseline fully connected neural network, e.g. the Multi-Layer Perceptron.

\parttitle{Conclusion}
Ph-CNN represents a novel deep learning approach for the classification of metagenomics data. Operatively, the algorithm has been implemented as a custom Keras layer taking care of passing to the following convolutional layer not only the data but also the ranked list of neighbourhood of each sample, thus mimicking the case of image data, transparently to the user.
\end{abstract}

\end{abstractbox}
\end{fmbox}
\end{frontmatter}

\section*{Background}
Biological data is often complex, heterogeneous and hard to interpret, thus a good testbed for Deep Learning (DL) techniques~\cite{ching17opportunities}.
The superiority of deep neural network approaches is acknowledged in a first group of biological and clinical tasks, with new results constantly flowing in in the literature~\cite{mamoshina16applications,chaudhary17deep,zacharaki17prediction}.
However, DL  is not yet a ''silver bullet'' in bioinformatics; indeed a number of issues are still limiting its potential in applications, including limited data availability, result interpretation and hyperparameters tuning~\cite{min16deep}.
In particular, DL approaches has so far failed in showing an advantage in metagenomics, either in terms of achieving better performance or detecting meaningful biomarkers.
This lack of significant results led Ditzler and coauthors ~\cite{ditzler15multilayer} to state that deep learning ''may not be suitable for metagenomic application''; nevertheless, novel promising attempts have recently appeared ~\cite{arangoargoty17deeparg,fang17gcoda}.
\begin{figure*}[!tb]
\centering\includegraphics[width=0.9\textwidth]{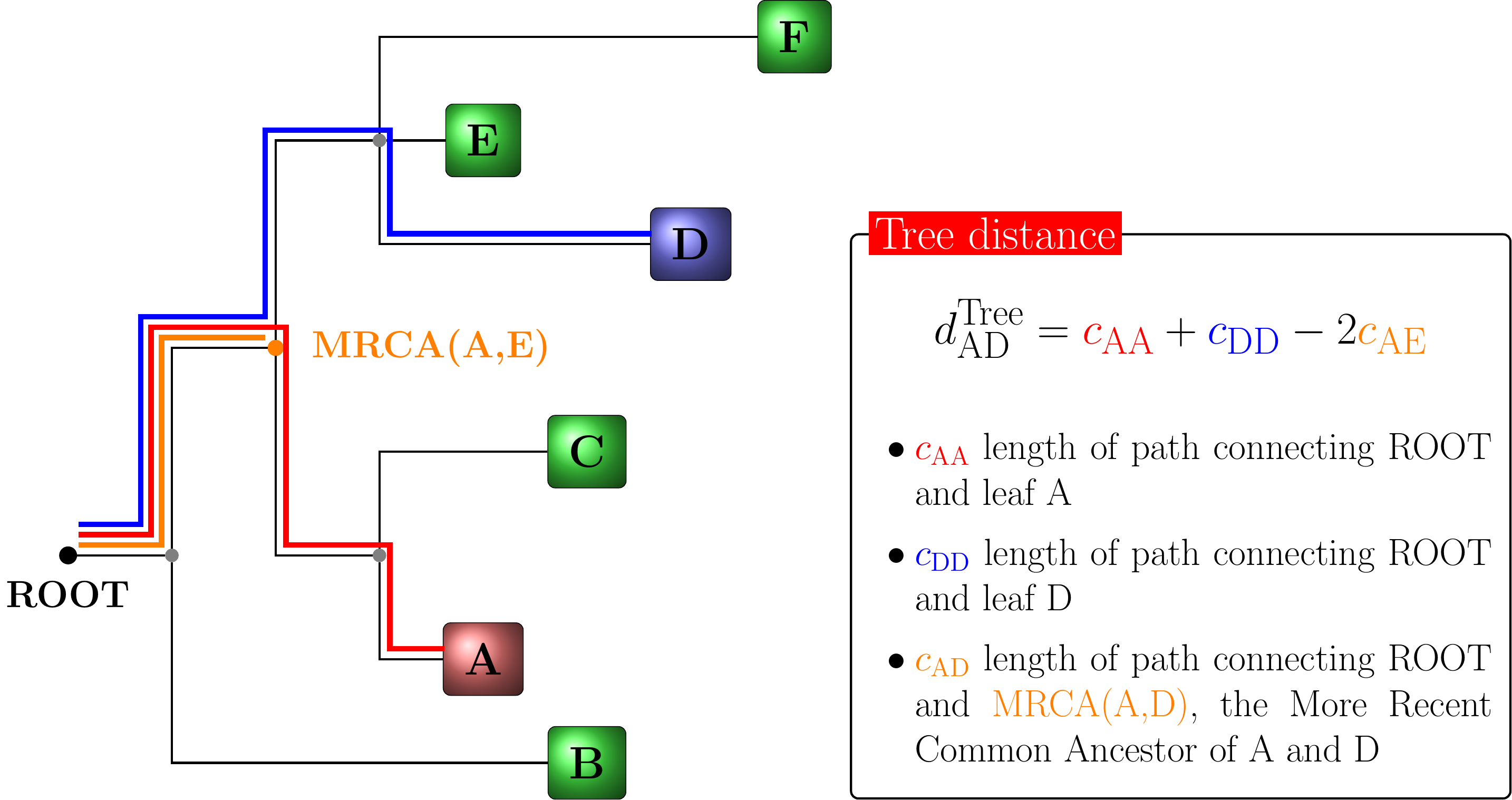}
\caption{\csentence{Patristic distance on a tree.}}
\label{fig:treedist}
\end{figure*}
Unique among other omics, metagenomics features are endowed with a hierarchical structure provided by the phylogenetic tree defining the bacterial clades.
In detail, samples are usually described by features called Operational Taxonomic Units (OTU). 
For each OT, its position as a leaf of the phylogenetic tree and its abundance value in the sample are automatically extracted by bioinformatics analysis.
In this work we exploit this hierarchical structure as an additional information  for the learning machine to better support the profiling process: this has been proposed before in~\cite{albanese15explaining,fukuyama17multidomain}, but only in shallow learning contexts, to support classification or for feature selection purposes.
We aim to exploit the phylogenetic structure to enable adopting the Convolutional Neural Network (CNN) DL architecture  otherwise not useful for omics data: we name this novel solution \emph{Ph-CNN}.
Indeed CNNs are the elective DL method for image classification~\cite{lecun98gradient,krizhevsky12imagenet} and they work by convolving subsets of the input image with different filters.
The operation is based on the matricial structure of a digital image and, in particular, the concept of neighbours of a given pixel.
Using the same architecture for non-image data requires the availability of an analogous proximity measure between features.
\begin{figure*}[!tb]
\centering\includegraphics[width=0.9\textwidth]{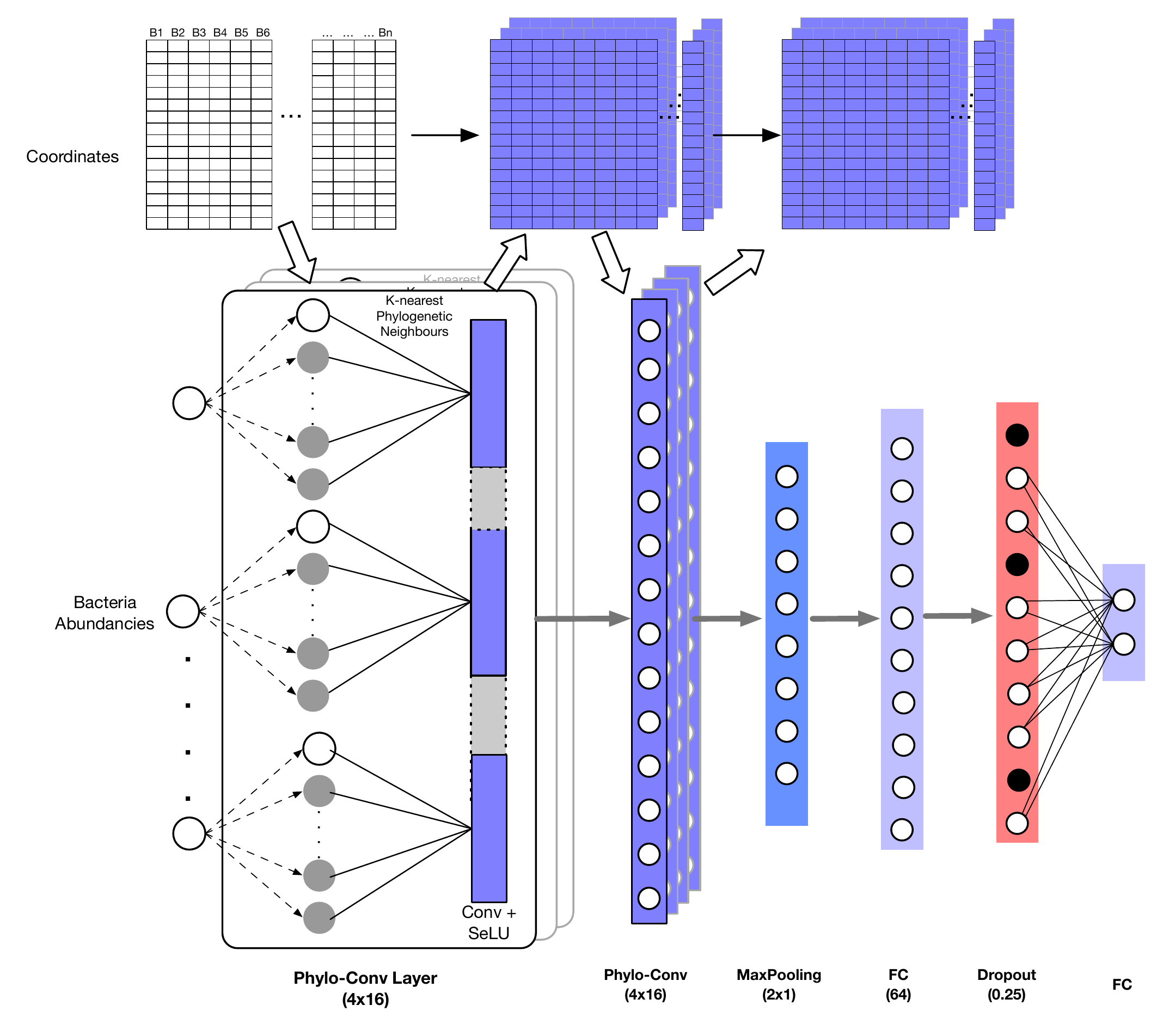}
\caption{\csentence{The structure of Ph-CNN.} In this configuration, Ph-CNN is composed by two PhyloConv layers followed by a Fully Connected layer before decision.}
\label{fig:phcnn}
\end{figure*}
In the metagenomics case, such measure can be inherited by the tree structure connecting the OTUs and the neighbourhood
 are naturally defined once an approprieate tree distance between two OTUs is defined.
In this paper, we adopt the patristic distance, \textit{i.e.}, the sum of the lengths of all branches connecting two OTUs on the phylogenetic tree~\cite{stuessy08patrocladistic}.
By definition, the output of a CNN consists of linear combinations of the original input features: this implies that, if Ph-CNN includes more CNN layers, the problem of finding the neighbours of a OTU is shifted into the hardest task of finding the neighbours of a linear combination of OTUs. 
The workaround here is mapping OTUs into points of a $k$-dimensional metric space preserving distances as well as possible via a MultiDimensional Scaling (MDS) projection~\cite{cox01multidimensional}: the use of MDS is allowed because the patristic distance is Euclidean~\cite{devienne11euclidean}.
A further refinement is provided by sparsifying MDS via regularized low rank matrix approximation~\cite{shen07sparse} through the addition of the smoothly clipped absolute deviation penalty~\cite{fan01variable}, tuned by cross-validation.

The convolutional layer combined with the neighbours detection algorithm is operatively implemented as a novel Keras layer~\cite{chollet15keras} called Phylo-Conv. 
Ph-CNN consists of a stack of Phylo-Conv layers first flattened then terminating with a Fully Connected (Dense) and a final classification layer.
The experimental setup is realized as a 10x5-fold cross-validation schema with a feature selection and ranking procedure, implementing the Data Analysis Protocol (DAP) developed within the US-FDA led initiatives MAQC/SEQC~\cite{maqc10maqc,seqc14comprehensive}, to control for selection bias and other overfitting effects and warranting honest perfomance estimates on external validation data subsets. 
Top ranking features are recursively selected as the $k$-best at each round, and finally aggregated via Borda algorithm~\cite{jurman12algebraic}.
Model performance is computed for increasing number of best ranking features by Matthews Correlation Coefficient (MCC), the measure that better convey in an unique value the confusion matrix of a classification task, even in the multiclass case~\cite{matthews75comparison,baldi00assessing,jurman12comparison}.
Experiments with randomized features and labels are also performed as model sanity check.

We demonstrate Ph-CNN characteristics with experiments on both synthetic and real omics data.
For the latter type, we consider  Sokol's lab data~\cite{sokol17fungal} of microbiome information for 38 healthy subjects (HS) and 222 inflammatory bowel disease (IBD) patients. 
The bacterial composition was analysed using 16S sequencing and a total number of 306 different OTUs was found.
IBD is a complex disease arising as a result of the interaction of environmental and genetic factors inducing immunological responses and inflammation in the intestine and primarily including ulcerative colitis (UC) and Crohn’s disease (CD).
Both disease classes are characterized by two conditions: flare (f), when symptoms reappear or worsen, and remission (r), when symptoms are reduced or disappear. 
Finally, since CD can affect different parts of the intestine, we distinguish ileal Crohn’s disease (iCD) and colon Crohn’s disease (cCD).
Note however,that the number of non zero features varies for the different rom tasks to task, (defined by disease, condition site) since some features may vanish on all samples of a class. 

Synthetic data are constructed mimicking the structure of the IBD dataset.
They are generated as compositional data from multivariate normal distributions with given covariances and means: in particular, to provide different complexity levels in the classification task, four different instances of data are generated with different ratios between class means.
On both data types, the Ph-CNN architecture than compared with state-of-art shallow algorithms as Support Vector Machines (SVMs) and Random Forest (RF), and with alternative neural networks methods such as Multi-Layer Perceptron (MLPNN).
\begin{table}[!bt]
\caption{Patient stratification in the IBD dataset.}
\label{tab:IBD_comp}
\begin{tabular}{|c|c|c|c|c|c|c|}
\hline
HS & \multicolumn{6}{c|}{IBD patients} \\ 
\hline
   & \multicolumn{2}{c|}{CDf} & \multicolumn{2}{c|}{CDr} & UCf & UCr \\ 
\cline{2-7}
  & iCDf & cCDf & iCDr & cCDr & &\\ 
\hline
38 & 44 & 16 & 59 & 18 & 41 & 44 \\
14.6\% & 16.9\% & 6.1\% & 22.7\% & 6.9\% & 15.8\% & 16.9\% \\ 
\hline
\end{tabular}
\end{table}
Moreover, the bacterial genera detected as top discriminating features are consistent with the key players known in the literature to play a major role during the IBD progression. Since the direct use of Ph-CNN on the IBD dataset leads to overfitting after few epochs due to the small sample size, the IBD dataset is used in a transfer learning (domain adaptation) task.

A preliminary version of the method has been presented as the M.Sc. thesis~\cite{giarratano16phylogenetic}.
\begin{figure*}[!tb]
\centering\includegraphics[width=0.9\textwidth]{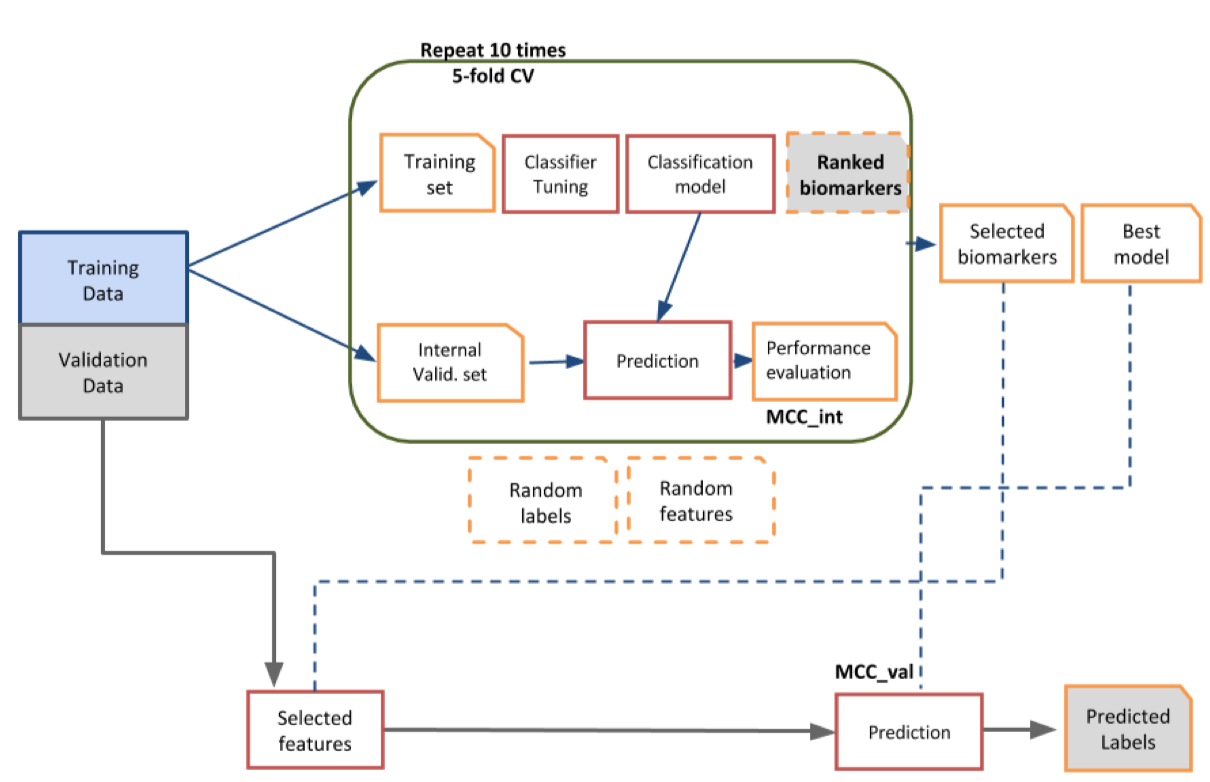}
\caption{\csentence{Data Analysis Protocol for the experimental framework.}}
\label{fig:dap}
\end{figure*}
\section*{Methods}

\subsection*{Ph-CNN}
The Ph-CNN is a novel DL architecture aimed at effectively including the phylogenetic structure of metagenomics data into the learning process.
The core of the network is the Phylo-Conv layer, a novel Keras~\cite{chollet15keras} layer coupling convolution with the neighbours detection.
In a generic Phylo-Conv layer, the structure input is represented by a collection of meta-leaves, \textit{i.e.} linear combinations of the leaves of the original tree; for the first Phylo-Conv layer, the structure input is simply the original set of leaves (OTUs, in the metagenomic case).
The neighbour detection procedure identifies the $k$-nearest neighbours of a given metaleaf to be convolved with the filters by the CNN.
The core ingredient is the choice of a metric on the phylogenetic tree~\cite{stjohn17review,entringer97distance} quantifying the distance between two leaves on the tree.
In the current case, we choose the patristic distance~\cite{stuessy08patrocladistic}, \textit{i.e.}, the sum of the lengths of all branches connecting two OTUs.
In Fig.~\ref{fig:treedist} we show how to compute the patristic distance between two leaves in a tree.

To deal with the problem of finding neighbours for linear combinations of leaves, we map the discrete space of the set of leaves into an Euclidean space of a priori chosen dimension, by associating each leaf to a point $P_i$ in the Euclidean space with variable Euclidean coordinates preserving the tree distance as well as possible.
The algorithm used for this mapping is the metric Multidimensional Scaling (MDS) \cite{cox01multidimensional}, whose use is allowed because the square root $\sqrt{d^\textrm{Tree}}$ of the patristic distance in Fig.~\ref{fig:treedist} is euclidean~\cite{devienne11euclidean}, that is, the matrix $(P_i\cdot P_j)$ is positive semidefinite.
\begin{figure*}[!tb]
\centering\includegraphics[width=0.9\textwidth]{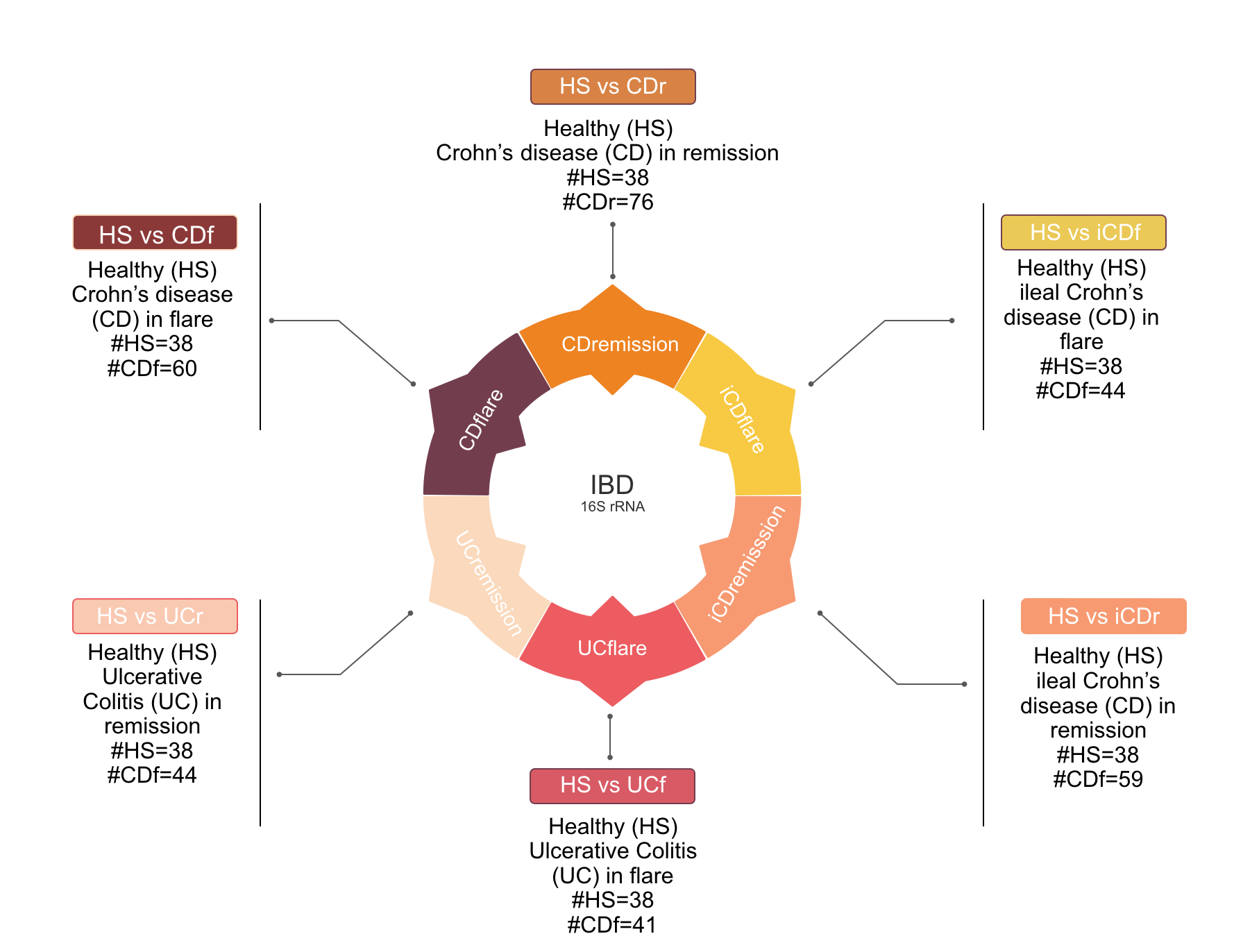}
\caption{\csentence{Classification tasks on IDB dataset.} The six learning tasks discriminating HS versus different stages of IBD patients.}
\label{fig:tasks}
\end{figure*}
Thus, given a linear combination of OTUs, it is possible to compute its $k$-nearest neighbours as the $k$-nearest neighbours of the corresponding linear combination of projected points $P_i$: in all experiments, the number of neighbours $k$ is set to 16.
The selected neighbours are then convolved with the 16 filters on the CNN.
The Phylo-Conv is then repeated; finally, the terminating layers of the Ph-CNN are a MaxPooling, then a Flatten layer and, finally, a Fully Connected with 64 neurons (changed to 128 for the transfer learning experiemnts) and a 0.25 Dropout.
Each convolutional layer has a Scaled Exponential Linear Units (SELU) \cite{klambauer17selfnormalizing} as the activation fuction,and the dense layer in transfer learning experiments uses a sigmoid activation function.
Adam~\cite{kingma14adam} is used as optimizer with learning rate 0.0005.
\begin{figure*}[!tb]
\centering\includegraphics[width=0.9\textwidth]{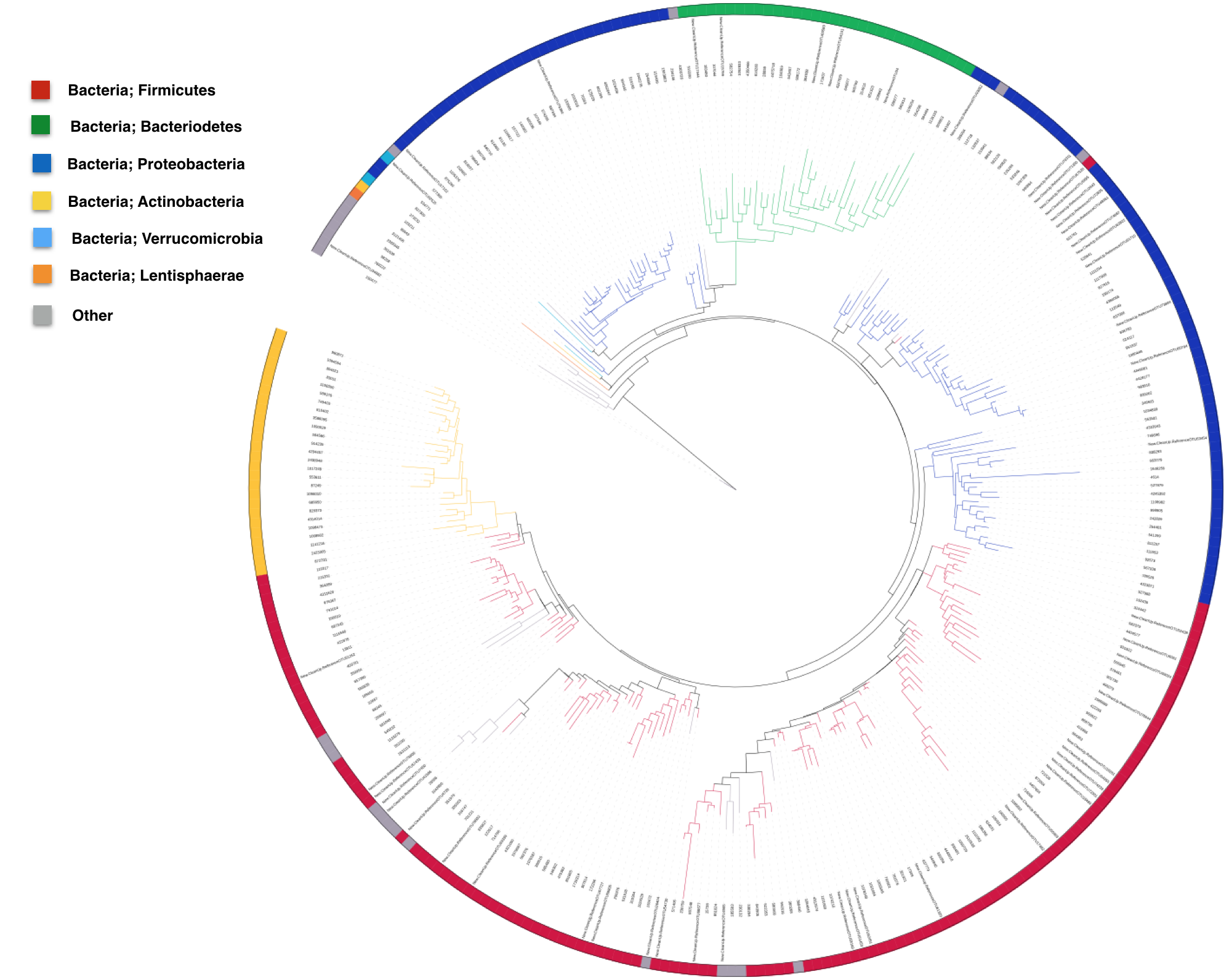}
\caption{\csentence{The phylogenetic tree for the IDB dataset.}}
\label{fig:tree}
\end{figure*}
\subsection*{Experimental setup}
To ensure predictive power and limit overfitting effect, the experimental framework is structured following the guidelines recommended by the US-FDA led studies MAQC/SEQC~\cite{maqc10maqc,seqc14comprehensive} that investigated the development of predictive models for the analysis of high-throughput data. 
In particular, the Ph-CNN becomes the core of an experimental setup designed according to the DAP shown in Fig.~\ref{fig:dap}, based on 10 repetitions of a 5-fold cross validation.

In details, the dataset is first partitioned into a non overlapping training set and test set, preserving the original stratification, \textit{i.e.}, the ratio between 
sample size across classes.
In the experiments described hereafter, the training set size is 80\% of the original dataset.
Then the training set undergoes 10 rounds of 5-fold stratified cross validation, with Ph-CNN as the classifier and $k$-Best as the feature selection algorithm, with ANOVA F-value as the ranking score.
At each round, several models are built for increasing number of ranked features (in this case, 25\%, 50\%, 75\% and 100\% of the total features) using Matthews Correlation Coefficient (MCC) \cite{matthews75comparison,baldi00assessing} as the performance measure.
MCC is rated as an elective choice ~\cite{maqc10maqc,seqc14comprehensive} for effectively combining into a single figure the confusion matrix of a classification task, and hence for evaluating classifiers' outcomes even when classes are imbalanced. Originally designed for binary discrimination, a multiclass version has also been developed~\cite{gorodkin04comparing,jurman12comparison}.
MCC values range between -1 and 1, where 1 indicates perfect classification, -1 perfect misclassification and 0 for coin tossing or attribution of every samples to the largest class.
The lists of ranked features produced within the cross-validation schema are then fused into a single ranked list using the Borda method~\cite{borda81memoire,saari01chaotic,jurman08algebraic}.
The subset of the fused list of ranked featured corresponding to the higher MCC value is selected as the optimal set of discriminating features for the classification tasks.
The fused list is further used to build the models for increasing number of features on the validation set (sometimes called the external validation set, to avoid ambiguities with the internal validation sets created at each CV round).
Finally, as sanity check for the procedure, the same methodology is applied several times on instances of the original dataset after randomly permuting the labels (random labels in Fig.~\ref{fig:dap}) and picking up random features instead of selecting them on the basis of the model performances (random features in Fig.~\ref{fig:dap}): in both cases, a procedure unaffected by systematic bias should return an average MCC close to 0.

\subsection*{The IBD dataset}
The IBD dataset has been originally published in~\cite{sokol17fungal} for a study aimed at investigating correlation between bacteria and fungal microbiota in different stages of Inflammatory Bowel Disease.
IBD is a clinical umbrella term defining a group of inflammatory conditions of the digestive tract, induced by the interactions of environmental and genetic factors leading to immunological responses and inflammation in the intestine: Ulcerative colitis (UC) and Crohn's disease (CD) are the two main conditions. 
The onset of bacterial dysbiosis of the gut microbiota has recently been observed in patients affected by IBD: a decrease in the abundance of \textit{Firmicutes} phylum and an increase for \textit{Proteobacteria} phylum, albeit the exact pathogenesis of IBD remains unknown~\cite{morgan12dysfunction,sokol08faecalibacterium}.

The IBD dataset includes both fungal and bacterial abundances from faecal samples of 38 healthy subjects (HS) and 222 IBD patient, collected at the Gastroenterology Department of the Saint Antoine Hospital (Paris, France).
In the present study, we only consider the bacterial data subset.

IBD patients are divided in two classes according to the disease phenotype UC and CD. 
Each disease class is further characterized by two conditions: flare (f), if symptoms reappear or worsen, and remission (r), if symptoms are reduced or disappear. 
Moreover, since CD can affected different parts of the intestine we further partition the data subset into ileal Crohn's disease (iCD) and colon Crohn's disease (cCD).
In Tab.~\ref{tab:IBD_comp} we summarize the sample distribution. 
In terms of learning tasks, we investigate the six classification tasks discriminating HS versus the six IBD partitions UCf, UCr, CDf, CDr, iCDf and iCDr, as graphically shown in Fig.~\ref{fig:tasks}.

\begin{figure*}[!tb]
\centering\includegraphics[width=0.9\textwidth]{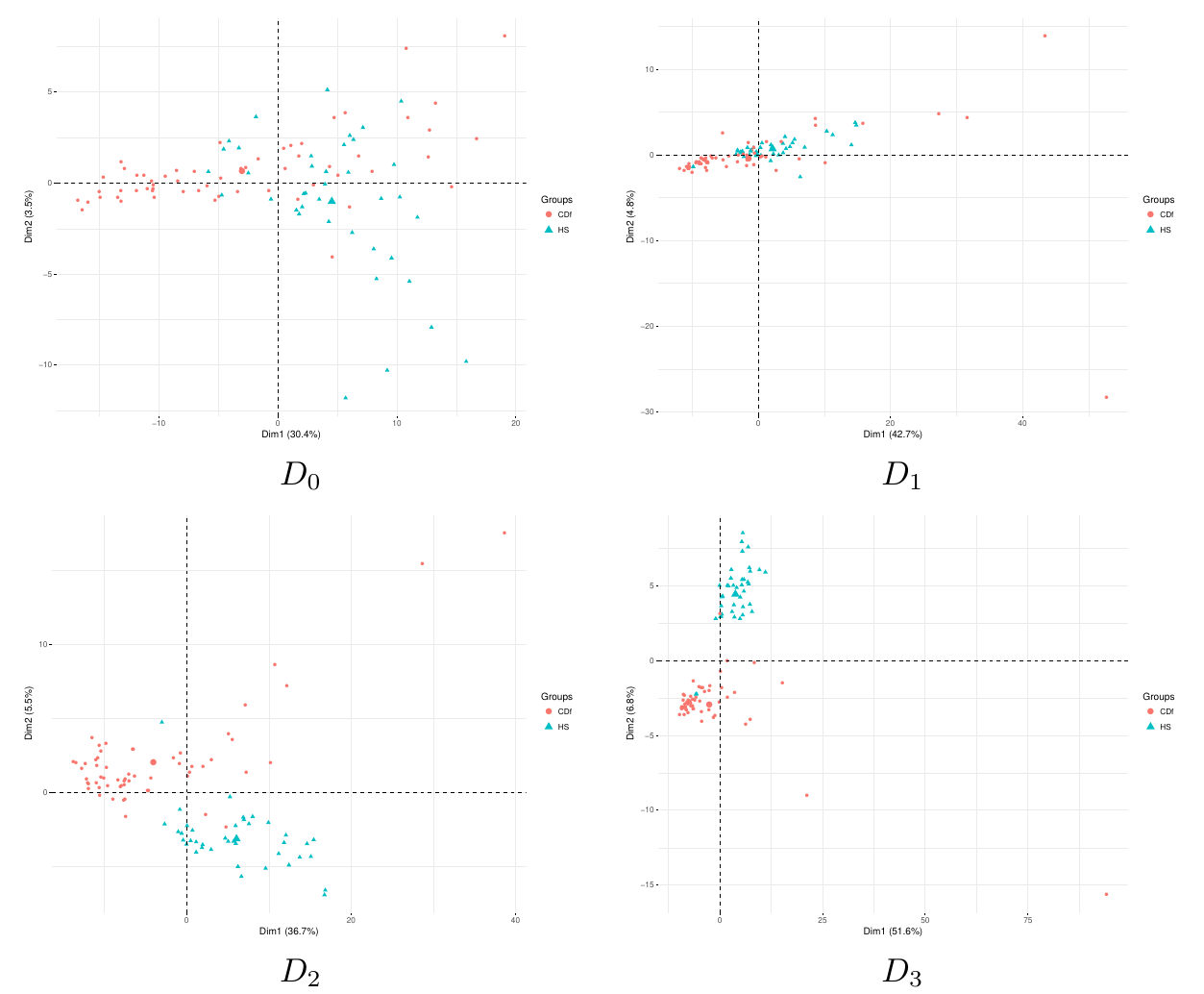} \\
\caption{\csentence{Principal component analysis for the 4 synthetic datasets $D_0,D_1,D_2,D_3$, with same sample sizes as in the IBD dataset.} Larger values of $\alpha$ correspond to more separate classes HR and CDf.}
\label{fig:pca}
\end{figure*}
The bacterial composition is analysed using 16S rRNA sequencing, demultiplexed and quality filtered using the QIIME 1.8.0 software~\cite{kuczynski05using,caporaso10qiime}; minimal sequence length was 200pb.
Sequences are assigned to OTUs using the UCLUST~\cite{edgar10search} algorithm with 97\% threshold pairwise identity and taxonomically classified using Greengenes reference database~\cite{mcdonald12improved}.
Samples with less than 10,000 sequences are excluded from analysis. 
The number of different OTUs found is 306: each OTU in the data sets is associated to the sequences with the same taxonomy. 
Among those sequences, the one with the highest median abundance across samples is chosen as the OTU representative. 
Since many sequences are not in the Greengenes database, OTUs can have an unassigned taxonomy: in this case, the OTU is removed from the analysis.
The actual number of OTUs used in the analyses is 259: for some discrimination tasks, however, the number of features is smaller, since some of them are all zeros for all samples in a class.
\begin{figure*}[!tb]
\centering\includegraphics[width=0.8\textwidth]{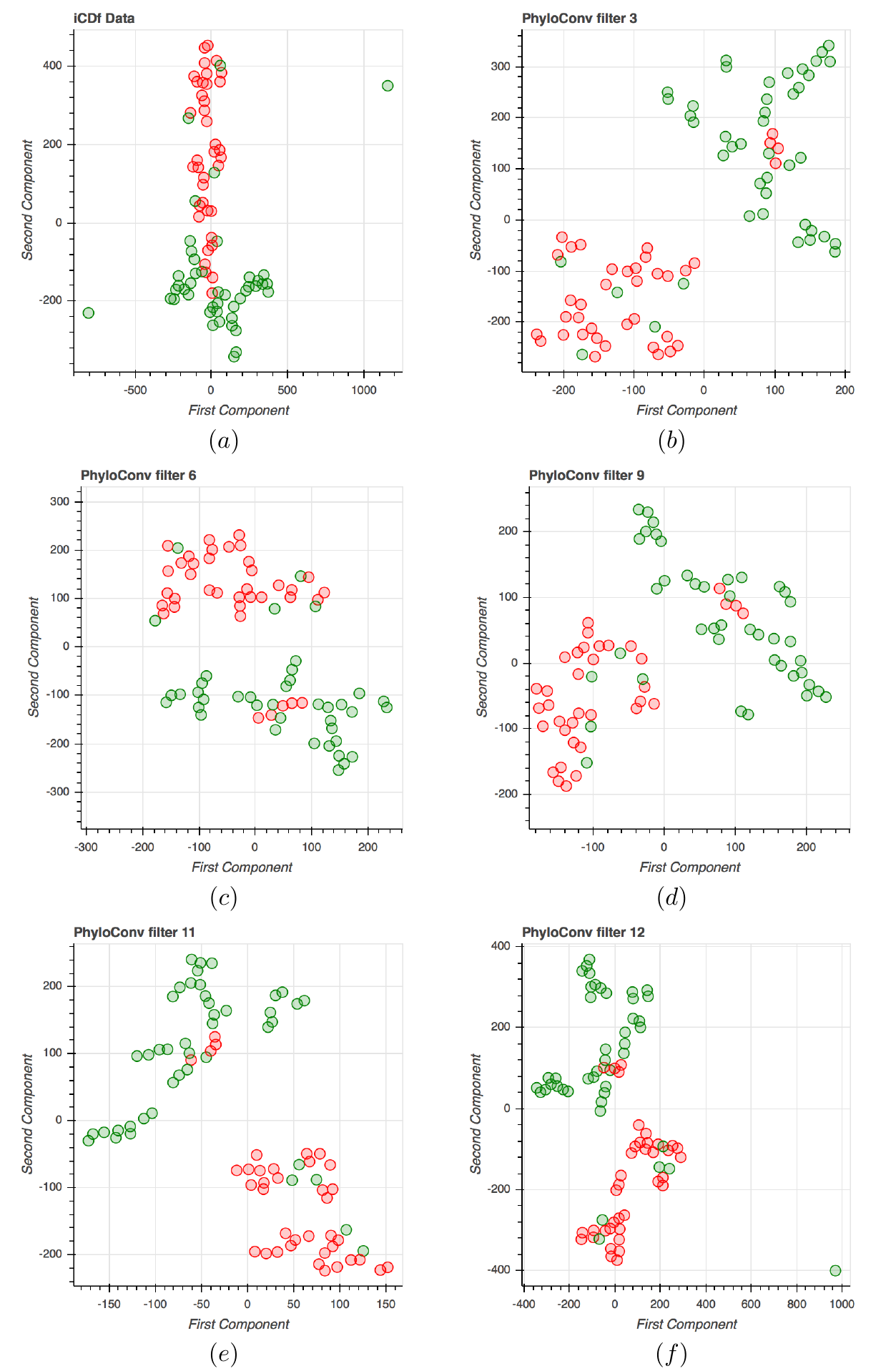} \\
\caption{\csentence{t-SNE projections of the original features} at initial layer (subfigure a) and after 3, 6, 9, 11, 12 convolutional filters (subfigures b-f). Green for healthy subjects, red for iCDf patients.}
\label{fig:tsne}
\end{figure*}
The distance between the OTUs is inferred first by aligning sequences using the NAST algorithm~\cite{desantis06nast,caporaso09pynast} and then by building the phylogenetic tree via the RAxML algorithm~\cite{stamatakis14raxml}.
The phylogenetic tree for the IBD dataset resulting from the described procedure is shown in Fig.~\ref{fig:tree}: largest abundance values of gut microbiota belong to \textit{Firmicutes} (red), \textit{Bacteroidetes} (green) and \textit{Proteobacteria} (blue), consistently with the published literature.
\begin{table*}[!tb]
\caption{Dataset $\textbf{D}$: classification performances of Ph-CNN compared to other classifiers on six classification tasks. Performance measure is MCC, with 95\% studentized bootstrap confidence intervals (min CI, max CI). Models are computed for $p= \{ 25\%, 50\%, 75\% \textrm{ and }100\%\}$ of total number of features for each task. Comparing algorithms are linear Support Vector Machines (LSVM), Random Forest (RF) and MultiLayer Perceptron (MLPNN).}
\label{tab:synth_int}
\begin{tabular}{lr|rrr|rrr|rrr|rrr}
 & & \multicolumn{3}{c|}{Ph-CNN} & \multicolumn{3}{c|}{LSVM} & \multicolumn{3}{c|}{MLPNN} & \multicolumn{3}{c}{RF}\\
\hline
Task & $p$ & MCC & min CI & max CI & MCC & min CI & max CI& MCC & min CI & max CI & MCC & min CI & max CI \\ 
\\
UCf &63 &0.794  &0.785  &0.803 &0.799 &0.793  &0.803 & 0.701  &0.692  &0.721 &0.729 &0.723  &0.736 \\
UCf &125  &0.852  &0.845  &0.860 &0.861 &0.857  &0.865 & 0.838  &0.834  &0.842 &0.843 &0.837  &0.849 \\
UCf &188  &0.920  &0.916  &0.925 &0.924 &0.921  &0.926 & 0.865  &0.861  &0.869 &0.902 &0.899  &0.906 \\
UCf &250  &0.940  &0.937  &0.944 &0.943 &0.941  &0.945 & 0.898  &0.894  &0.901 &0.903 &0.900  &0.907 \\
\\
UCr &  60 &0.861  &0.855  &0.867 &0.811 &0.807  &0.815 & 0.873  &0.869  &0.443 &0.797 &0.792  &0.801 \\
UCr & 119 &0.893  &0.888  &0.899 &0.866   &0.862  &0.870&0.877  &0.873  &0.877 &0.799 &0.794  &0.803 \\
UCr & 178  &0.906  &0.900  &0.911 &0.892 &0.888  &0.895 & 0.859 &0.855  &0.880 &0.791 &0.787  &0.794 \\
UCr & 237  &0.920  &0.916  &0.924 &0.917 &0.914  &0.920 &0.849  &0.844  &0.854 &0.790 &0.786  &0.795 \\
\\
CDf & 65  &0.785  &0.775  &0.795 &0.781 &0.776  &0.785&0.604  &0.593  &0.614 &0.764 &0.760  &0.769 \\
CDf & 130 &0.832  &0.825  &0.840 &0.833 &0.829  &0.838 & 0.821  &0.817  &0.825 &0.805 &0.800  &0.810 \\
CDf & 195 &0.896  &0.891  &0.901 &0.910 &0.907  &0.912 & 0.830  &0.825  &0.836 &0.863 &0.860  &0.867 \\
CDf & 259 &0.927  &0.924  &0.930 &0.920 &0.918  &0.923 &0.858 &0.854  &0.862 &0.880 &0.877  &0.883 \\
\\
CDr & 65  &0.714  &0.705  &0.723 &0.740 &0.734  &0.746 &0.498 &0.473  &0.521 &0.688 &0.682  &0.695 \\
CDr & 129 &0.799  &0.793  &0.806 &0.802 &0.798  &0.808 &0.783 &0.778  &0.788 &0.744 &0.740  &0.784 \\
CDr & 193 &0.850  &0.844  &0.856 &0.860 &0.857  &0.864 &0.766 &0.759  &0.773 &0.762 &0.756  &0.767 \\
CDr & 257 &0.890  &0.884  &0.895 &0.880 &0.877  &0.882 &0.788 &0.782  &0.794 &0.765 &0.761  &0.771 \\
\\
iCDf & 62 &0.781  &0.772  &0.790 &0.804 &0.799  &0.808   &0.845 &0.840  &0.849 &0.748 &0.743  &0.753 \\
iCDf & 124  &0.863  &0.854  &0.871 &0.861 &0.858  &0.865 &0.889 &0.886  &0.893 &0.808 &0.803  &0.814 \\
iCDf & 186  &0.922  &0.918  &0.926 &0.921 &0.919  &0.924 &0.879  &0.875  &0.883 &0.880 &0.877  &0.883 \\
iCDf & 247  &0.944  &0.941  &0.947 &0.941 &0.939  &0.942 &0.901 &0.899  &0.904 &0.890 &0.887  &0.893 \\
\\
iCDr & 65   &0.753  &0.744  &0.763 &0.773 &0.769  &0.779 &0.807  &0.802  &0.812 &0.724 &0.719  &0.729 \\
iCDr & 129  &0.830  &0.823  &0.837 &0.834 &0.830  &0.837 &0.822 &0.816  &0.827 &0.794 &0.788  &0.800 \\
iCDr & 193  &0.884  &0.878  &0.889 &0.893 &0.891  &0.896 &0.831 &0.827  &0.835 &0.812 &0.807  &0.818 \\
iCDr & 257  &0.910  &0.905  &0.915 &0.907 &0.904  &0.909 &0.837 &0.831  &0.842 &0.820 &0.816  &0.825 \\
\end{tabular}
\end{table*}
\subsection*{The synthetic datasets}
The synthetic datasets are generated as compositional data, \textit{i.e.}, vectors lying in the $p$-dim Aitchison simplex 
$\mathcal{S}=\left\{\mathbf{x}=\left( x_1, x_2, \dotsc, x_p \right) \in (\mathbb{R}_0^+)^p\;\textrm{with}
;\sum_{j=1}^{p} x_j = 1\right\}$, whose structure resembles the IBD data.

Note that the application of standard multivariate statistical procedures on compositional data requires adopting adequate invertible transformation procedures to preserve the constant sum constrain~\cite{aitchison86statistical}: a standard map is the isometric log ratio \textrm{ilr} \cite{egozcue03isometric}, invertibly projecting the $p$-dimensional Aitchison simplex isometrically to a $p-1$-dimensional euclidian vector.
Transformations like \textrm{ilr} allow using unconstrained statistics on the transformed data, with inferences mapped back to original compositional data through the inverse map.
\begin{table}[!bt]
\caption{Dataset $\textbf{D}$: classification performances of Ph-CNN compared to other classifiers on the external validation dataset.}
\label{tab:ext_s}
\begin{tabular}{c|cccc}
Task & Ph-CNN & LSVM & MLPNN & RF \\
\hline
UCf & 0.946 & 0.934 & 0.898 & 0.869\\
UCr & 0.897 & 0.904 & 0.897 & 0.756\\
CDf & 0.926 & 0.935 & 0.884 & 0.859\\
CDr & 0.888 & 0.888 & 0.821 & 0.722\\
iCDf & 0.931& 0.943 & 0.905 & 0.863\\
iCDr & 0.901& 0.910 & 0.846 & 0.778\\
\end{tabular}
\end{table}
The construction of the synthetic data starts from the IDB dataset, and in particular from the two subsets of the HS and CDf samples (by abuse of notation, we use the same identifier for both the class and the compositional data subset). 
Classes HS and CDf are defined by 259 features (OTU), and they include 38 and 60 samples respectively.
The key step is the generation of the synthetic $\textrm{HS}^\alpha_s$ and $\textrm{CDf}^\alpha_s$ subsets, sampled from multivariate normal distributions with given covariances and mean. 
\begin{table*}[!tb]
\caption{Dataset IBD: classification performances of Ph-CNN (in transfer learning from $\textbf{D}$) compared to other classifiers on six classification tasks. Performance measure is MCC, with 95\% studentized bootstrap confidence intervals (min CI, max CI). Models are computed for $p= \{ 25\%, 50\%, 75\% \textrm{ and }100\%\}$ of total number of features for each task. Comparing algorithms are linear Support Vector Machines (LSVM), Random Forest (RF) and MultiLayer Perceptron (MLPNN).}
\label{tab:idb_int}
\begin{tabular}{lr|rrr|rrr|rrr|rrr}
 & & \multicolumn{3}{c|}{Ph-CNN} & \multicolumn{3}{c|}{LSVM} & \multicolumn{3}{c|}{MLPNN} & \multicolumn{3}{c}{RF}\\
\hline
Task & $p$ & MCC & min CI & max CI & MCC & min CI & max CI& MCC & min CI & max CI & MCC & min CI & max CI \\
\\
UCf &63  &0.659  &0.604  &0.709 &0.510  &0.449 &0.573&0.689 &0.629 &0.743 &0.741 &0.698 &0.783 \\
UCf &125 &0.668  &0.595  &0.734 &0.438  &0.368 &0.500&0.644 &0.582 &0.703 &0.742 &0.690 &0.792 \\
UCf &188  &0.650  &0.599  &0.707 &0.541  &0.438 &0.604 &0.570 &0.496 &0.644 &0.735 &0.680 &0.784 \\
UCf &250  &0.628  &0.567  &0.687 &0.565  &0.510 &0.619 &0.606 &0.547 &0.667 &0.760 &0.707 &0.816 \\
\\
UCr &  60  &0.445  &0.375  &0.517 &0.509  &0.221 &0.384 &0.415 &0.350 &0.476 &0.508 &0.425 &0.584\\
UCr & 119  &0.464  &0.393  &0.537 &0.533  &0.238 &0.357&0.528 &0.463 &0.596 &0.455 &0.387 &0.525\\
UCr & 178  &0.444  &0.372  &0.520 &0.519  &0.328 &0.449&0.538 &0.471 &0.610 &0.435 &0.363 &0.504\\
UCr & 237  &0.346  &0.283  &0.536 &0.408  &0.303 &0.420&0.489 &0.417 &0.557 &0.400 &0.337 &0.463\\
\\
CDf & 65  &0.613  &0.555  &0.665 &0.419  &0.363 &0.472&0.610 &0.549 &0.666 &0.677 &0.618 &0.728 \\
CDf & 130 &0.617  &0.549  &0.601 &0.326  &0.252 &0.394&0.620 &0.551 &0.685 &0.706 &0.648 &0.758 \\
CDf & 195  &0.630  &0.560  &0.682 &0.647  &0.595 &0.691&0.601 &0.534 &0.667 &0.739 &0.685 &0.788 \\
CDf & 259 &0.572  &0.501  &0.620 &0.595  &0.545 &0.642 &0.648 &0.589 &0.703 &0.720 &0.667 &0.768 \\
\\
CDr & 65  &0.241  &0.172  &0.311 &0.138  &0.073 &0.198&0.235 &0. &0.306 &0.488 &0.437 &0.541\\
CDr & 129 &0.232  &0.167  &0.295 &0.089  &0.028 &0.151 &0.275 &0.199 &0.348 &0.432 &0.373 &0.485 \\
CDr & 193 &0.202  &0.131  &0.273 &0.169  &0.101 &0.236  &0.243 &0.172 &0.315 &0.402 &0.341 &0.464\\
CDr & 257 &0.218  &0.158  &0.278 &0.178  &0.107 &0.251  &0.233 &0.160 &0.305 &0.398 &0.331 &0.464  \\
\\
iCDf & 62 &0.704  &0.655  &0.753 &0.534  &0.484 &0.583&0.679 &0.622 &0.739 &0.787 &0.746 &0.831 \\
iCDf & 124  &0.702  &0.642  &0.760 &0.414  &0.346 &0.482  &0.690 &0.634 &0.743 &0.811 &0.766 &0.854\\
iCDf & 186  &0.680  &0.614  &0.738 &0.662  &0.605 &0.718 &0.685 &0.630 &0.742 &0.791 &0.741 &0.836\\
iCDf & 247  &0.681  &0.614  &0.739 &0.561  &0.507 &0.621  &0.708 &0.652 &0.764 &0.775 &0.730 &0.820 \\
\\
iCDr & 65  &0.537  &0.480  &0.601 &0.338  &0.277 &0.409  &0.526 &0.475 &0.581 &0.552 &0.492 &0.612 \\
iCDr & 129  &0.522  &0.453  &0.595 &0.319  &0.254 &0.385  &0.558 &0.493 &0.623 &0.563 &0.516 &0.609\\
iCDr & 193  &0.556  &0.492  &0.617 &0.377  &0.315 &0.437 &0.459 &0.388 &0.527 &0.566 &0.516 &0.616 \\
iCDr & 257  &0.477  &0.411  &0.544 &0.438  &0.378 &0.492 &0.529 &0.462 &0.598 &0.539 &0.482 &0.596 \\
\end{tabular}
\end{table*}
Operatively, let $\textrm{HS}'$ and $\textrm{CDf}'$ the ilr-transformed $\textrm{HS}$ and $\textrm{CDf}$ subsets.
Then compute the featurewise mean $\mu(\textrm{HS}')=\left(\mu_1(\textrm{HS}'),\mu_2(\textrm{HS}'),\ldots,\mu_{258}(\textrm{HS}')\right)$ and $\Sigma(\textrm{HS}')$ the covariance matrix.
Analogously compute $\mu(\textrm{CDf}')$ and $\Sigma(\textrm{CDf}')$.
Consider now the matrix $\textrm{HS}'_0$ defined by substracting to each row of $\textrm{HS}'$ the vector of the means: $\left(\textrm{HS}'_0\right)_{i\cdot} = (\textrm{HS}')_{i\cdot}-\mu(\textrm{HS}')$, and define analogousy the matrix $\textrm{CDf}'_0$ by $\left(\textrm{CDf}'_0\right)_{i\cdot} = (\textrm{CDf}')_{i\cdot}-\mu(\textrm{HS}')$.
\begin{table}[!tb]
\caption{Dataset $\textbf{D}$ on IBD: classification performances of Ph-CNN compared to other classifiers on the external validation dataset.}
\label{tab:ext}
\begin{tabular}{c|cccc}
Task & Ph-CNN & LSVM & MLPNN & RF \\
\hline
UCf & 0.741 & 0.740 & 0.666 & 0.699\\
UCr & 0.583 & 0.497 & 0.608 & 0.678\\
CDf & 0.858 & 0.642 & 0.705 & 0.707 \\
CDr & 0.853 & 0.654 & 0.654 & 0.597 \\
iCDf& 0.842 & 0.418 & 0.401 & 0.920 \\
iCDr& 0.628 & 0.414 & 0.414 & 0.418 \\
\end{tabular}
\end{table}
Introduce the projections $P_{\textrm{HS}'}=\textrm{HS}'_0\cdot (\mu(\textrm{HS}')-\mu(\textrm{CDf}'))$ and $P_{\textrm{CDf}'}=\textrm{CDf}'_0\cdot (\mu(\textrm{HS}')-\mu(\textrm{CDf}'))$, then define now 
$\sigma=\sqrt{ \frac{\sum_{i=1}^{38} (P_{\textrm{HS}'})_i^2 + \sum_{i=1}^{60} ( (P_{\textrm{CDf}})_i - (\mu_i(\textrm{CDf}')-\mu_i(\textrm{HS}')))^2}   {38+60}}$ and 
$\mu= \frac{\mu(\textrm{HS}')+\mu(\textrm{CDf}')}{2}$.
Fix $\alpha\in\mathbb{R}_0^+$ and define $m_\textrm{HS}=\mu+\alpha\sigma\frac{\mu(\textrm{HS}')}{||\mu(\textrm{HS}')||}$ and 
$m_\textrm{CDf}=\mu+\alpha\sigma\frac{\mu(\textrm{CDf}')}{||\mu(\textrm{CDf}')||}$.
Then, define $\textrm{HS'}^\alpha_s$ as the dataset collecting $n_\textrm{HS}$ instances from a multivariate normal distribution with mean $m_\textrm{HS}$ and covariance $\Sigma(\textrm{HS}')$ and analogously $\textrm{CDf'}^\alpha_s$.
The two synthetic data subsets $\textrm{HS}^\alpha_s$ and $\textrm{CDf}^\alpha_s$ are defined by taking ilr-counterimages:
$\textrm{HS}^\alpha_s = \textrm{ilr}^{-1}(\textrm{HS'}^\alpha_s)$ and $\textrm{CDf}^\alpha_s = \textrm{ilr}^{-1}(\textrm{CDf'}^\alpha_s)$.
Finally, the synthetic dataset $D_\alpha$ is then obtained as the union $\textrm{HS}^\alpha_s\cup\textrm{CDf}^\alpha_s$.
Setting the parameter $\alpha$, we provide different complexity levels in the classification task.
For instance, for $\alpha =0 $ the means of the two classes in the synthetic dataset $D_0$ are the same, while for $\alpha=1$ the means of the two classes HS and CDf are the same as in the IBD dataset; larger values of $\alpha$ correspond to easier classification tasks.
Principal component analysis of the four datasets $D_0, D_1, D_2, D_3$ with same sample size as IBD dataset is displayed in Fig.~\ref{fig:pca}.
With the same procedure, a synthetic dataset $\textbf{D}$ is created with 10,000 samples and $\alpha=1$, preserving class size ratios.
In practice, generation of the synthetic datasets was performed using the R packages \textit{compositions} \url{https://cran.r-project.org/web/packages/compositions/} \cite{denboogart08compositions} and \textit{mvtnorm} \url{https://cran.r-project.org/web/packages/mvtnorm/index.html} \cite{mi09mvtnorm}.

\section*{Results and discussion}
The $10\times 5-$fold CV DAP has been applied on instances of the synthetic datasets and on the IBD datasets, comparing the performance with standard (and shallow)  learning algorithms such as linear Support Vector Machines (SVM) and Random Forest (RF), and with a standard Multi Layer Perceptron (MLPNN)~\cite{bishop95neural}.
As expected~\cite{angermueller16deep}, no classification task can be reliably tackled by Ph-CNN using the IBD dataset alone: the very small sample size causes the neural network to overfit after just a couple of epochs.
To overcome this issue we explore the potentialities of transfer learning, as described in what follows.

As a first experiment, we apply the DAP on $\textbf{D}$.
In this case, the SELU activation function is used for every layer.
The results of the Ph-CNN DAP on $\textbf{D}$ are listed in Tab.~\ref{tab:synth_int} (internal validation) and Tab.~\ref{tab:ext_s} (external validation) on the six classification tasks Healthy vs. \{UCf, UCr, CDf, CDr, iCDf and iCDr\}; MCC on DAP internal validation is shown with 95\% studentized bootstrap confidence intervals~\cite{diciccio96bootstrap}.

The second experiment instead is based on a domain adaptation strategy.
The Ph-CNN is first trained on the synthetic dataset $\textbf{D}$, then all layer but the last one are freezed, the last layer is substituted by a 2-neurons Dense layer and then retrained on the IBD dataset.
Since only the last layer is trained in the second step, the term domain adaptation is best describing the methodology rather than the more generic transfer learning.
Here, the activation function is the ReLU for every layer.
The results of the Ph-CNN DAP together with the comparing classifiers are listed in Tab.~\ref{tab:ibd_int} (internal validation) and Tab.~\ref{tab:ext} (external validation).

As an observation, Ph-CNN tends to misclassify more the samples in class Healthy, rather than those in the other class, for each classification task.
In Fig.~\ref{fig:tsne} we show the embeddings of the original features at 6 different levels (after initial input and after 5 PhyloConv filters) for the iCDf task (IBD dataset) by projecting them in two dimensions via t-distributed Stochastic Neighbor Embedding (t-SNE) \cite{tsne2008} with perplexity = 5 and 5,000 iterations. While at input level the problem seems hardly separable, the classes tend to form distinct clusters during the flow through convolutional filters applied on OTUs close in the taxonomy. 

\paragraph{Computational details} The Ph-CNN is implemented as a Keras v2.0.8 layer, while the whole DAP is written in Python/Scikit-Learn~\cite{pedregosa11scikit}.
All computations were run on a Microsoft Azure platform with 2x {NVIDIA} Tesla K80 GPUs.

\section*{Conclusions}
We introduced here Ph-CNN, a novel DL approach for the classification of metagenomics data exploiting the hierarchical structure of the OTUs inherited by the corresponding phylogenetic tree.
In particular, the tree structure is used throughout the prediction phase to define the concept of OTU neighbours, used in the convolution process by the CNN. 
Results are promising, both in terms of learning performance and biomarkers detection
Extensions of the Ph-CNN architecture are addressing the testing of different tree distances, optimization of  neighbours detection and of  the number of Phylo-Conv layers. Further,  different feature selection algorithms, either generic or DL-specific can be adopted ~\cite{nezhada17safs,roy15feature, li16deep}. Improvements are expected on the transfer learning and domain adaptation procedures, such as learning on synthetic data and testing on metagenomics, and applying to larger datasets. Finally, beyond the metagenomics applications, we observe that Ph-CNN is  general purpose algorithm, whose use can be extended to other structured data, e.g. transcriptomic data and the genomic distance, or population genetics with genetic distance, to name a few

\subsection*{Abbreviations}
CD: Crohn’s disease,
CNN: Convolutional Neural Networks,
DAP: Data Analysis Protocol,
IBD: Inflammatory Bowel Disease,
MCC: Matthews Correlation Coefficient,
MDS: MultiDimensional Scaling,
MLPNN: Multi-Layer Perceptron,
OTU: Operational Taxonomic Unit,
RF: Random Forest,
SVM: Support Vector Machine,
UC: Ulcerative Colitis

\begin{backmatter}

\section*{Availability of data and materials}
Data and code are available at \url{https://gitlab.fbk.eu/MPBA/phylogenetic-cnn}.

\section*{Acknowledgements}
  The authors wish to thank Alessandro Zandon{\`a} for help during the analyses of the IBD dataset and Calogero Zarbo for help in a first implementation of the approach.

\section*{Author's contributions}
  YG, VM and DF implemented and performed all computational work; MC supervised the biological aspects of the projects; CA supervised the mathematical aspects of the project; CA and GJ conceived the project; CF and GJ directed the project and drafted the manuscript. All authors discussed the results, read and approved the final manuscript.  

\bibliographystyle{vancouver} 
\bibliography{fioravanti17phcnn} 
\end{backmatter}
\end{document}